\documentstyle[12pt]{article}
\begin{document}

\begin{center}

{\Large \bf New and Improved\\
Superstring Phenomenology\\}
\vspace{.25in}
{\large Joseph D. Lykken\\}
\vspace{.25in}
{\it Theoretical Physics Dept.\\
Fermi National Accelerator Laboratory\\
P.O. Box 500, Batavia, IL 60510}
\end{center}

\begin{abstract}
Recent developments in string theory have important implications for
cosmology. Topics discussed here are inflation, the cosmological constant,
smoothing of cosmological singularities, and dark matter from parallel
universes. Talk presented at the International Workshop on Particle Physics
and the Early Universe (COSMO-98), 15-20 Nov, Asilomar, Monterey, CA.
\end{abstract}

\section*{Lost in (Moduli) Space}

The duality revolution which occurred during the past five
years has enormously advanced our knowledge and perspective regarding
the theory formerly known as superstrings. There is now overwhelming
evidence\cite{jhs98}
to the effect that there is a unique string theory (often
called M theory), which contains, in addition to strings, a variety
of membranes and even particle-like objects. In a kind of Planckian
democracy, none of these degrees of freedom can truely be regarded as
more fundamental than the others. In fact in one limit of this theory
(somewhat confusingly called the M theory limit), the string degrees of
freedom are completely absent. This recent explosion of our understanding
of nonperturbative string physics nicely complements the impressive
edifice of perturbative string knowledge built up during the previous
decade. All of which leads to the obvious question:

\begin{itemize}  
\item {\large If we know so much about string theory, why can't we predict
anything?}
\end{itemize}

The main reason for this embarrassing irony is that string theory does
not have a unique (consistent, stable,) ground state. This was known
to be true at the perturbative level for many years, but only recently
have we realized that this disturbing property appears to persist even
when we bring to bear our full arsenal of nonperturbative string dynamics.
At low energies string theory is described (mostly, at least,) by an
effective field theory; without a definite choice for the string vacuum
we cannot even specify the degrees of freedom of this low energy theory,
let alone the form or parameters of the effective lagrangian.

The problem is that we are lost in moduli space. The effective field
theory limit of string theory contains a number of scalar fields,
called moduli, with flat potentials. This is not surprising since
most known string vacua preserve some spacetime supersymmetry, and
moduli are a generic feature of supersymmetric field theories. We can
parametrize a moduli space by the vacuum expectation values (vevs)
of these scalar fields. As we move around in this moduli space, the
effective field theory, defined by shifting by these vevs, can vary
enormously. Not only are there variations in couplings, but at special
points in moduli space the number and type of light degrees of freedom
changes.

We have only just begun to feel our way around the intricate tapestry
which represents the full moduli space of consistent string vacua.
We have probed around the edges, which represent the various possible
perturbative limits of string theory, as well as the 11-dimensional
(non-stringy) M theory limit. The interior remains largely
{\it terra incognita}, although string duality relations help us
trace the threads connecting these different limits. We know neither
the dimensionality nor the topology of the full string moduli space.
Nor do we know the connectivity of this space. The tapestry may be
very frayed, with many ragged patches connected to the main body by
only a few threads; there may even be ``string islands'',
points or patches of moduli space completely disconnected from
the main body.

One (at least) of these points ought to correspond to the Standard Model
particle physics and FRW cosmology that we observe at low
energies and long distance scales. But where?

\section*{Moduli and Cosmology}

It is not suprising that the existence of moduli fields
has implications for cosmology. Indeed this is true even
for approximate moduli, i.e., scalars whose flat potentials
are lifted by nonperturbative effects. Since the number and
type of moduli vary as we move around in moduli
space, most of what we can say about string cosmology is
very dependent on where we think the string ground state is.

Certain moduli, however, have slightly more robust characteristics.
Perturbative limits of string theory contain a dilaton, a weakly
coupled scalar whose vev determines the string coupling, and thereby
the basic relationships between the Planck scale, compactification
scales, and gauge couplings. These same perturbative limits also
contain a pseudoscalar axion, and indeed axions seem to be generic
features of large classes of string vacua. Compactified dimensions
in string theory can assume a wide variety of geometries;
nevertheless certain features of the modulus describing the ``overall''
scale of compactification are somewhat generic.

Another interesting class of moduli for cosmology are what I
will call ``invisible'' moduli. The vevs of these scalars describe
either dimensionless couplings or new mass and length scales associated
with hidden sectors- exotic matter and gauge fields which couple only
gravitationally to ordinary matter.

The good news for cosmologists is that string moduli provide
natural candidates for the scalar fields that may perform some
cosmologically important tasks. These include the inflaton and
perhaps quintessence.
The bad news for cosmologists is that generic regions of
string moduli space won't look good cosmologically. Indeed
generically moduli are more likely to be a cosmological headache
than a panacea. The devil, furthermore, is in the details, and
generically these details are difficult to tackle.    

The dilaton, for example, has properties which make
it an attractive candidate for 
the inflaton\cite{binetruy86,cvetic92,banks95}.
The dilaton acquires a nonvanishing potential only from 
nonperturbative effects; the relative gradient in this potential is
naturally of order the inverse Planck mass, as desired for slow roll
inflation.
On the other hand, there are a number of problems with
dilaton inflation\cite{campbell91,brustein93,banks96}.
Specific scenarios require additional assumptions about the
nonperturbative contributions to the potential; these assumptions
are hard to pin down with our current level of knowledge. 
In many scenarios there is the additional problem that the
dilaton kinetic energy dominates the potential energy. 
Furthermore, near any of
the perturbative limits of string theory,
the dilaton is generically unstable; it's vev wants to run off
to infinity, producing an infinitely weakly coupled theory.
One can postulate nonperturbative fixes for this runaway behavior,
but such scenarios are neither rigorous nor robust.

\section*{String Islands and the Cosmological Constant}

The cosmological constant problem is the most notorious and
vexing problem of quantum gravity\cite{cohn98}. 
Any attempt to unify quantum
mechanics with gravity leads (at least naively) to the conclusion
that quantum fluctuations in the vacuum (i.e. zero-point energies)
must couple to gravity. Since these zero-point energy sums are
typically divergent, their natural scale in a quantum field theory
is some ultraviolet cutoff $U$. One thus expects to generate a
cosmological constant of order
\begin{equation}
\Lambda \sim U^4 \quad ,
\end{equation}
and that the entropy associated with a system of linear size $L$
scales like
\begin{equation}
S \sim L^3U^3 \quad .
\end{equation}
Note that $\Lambda$ is positive if bosonic modes dominate the sum,
and negative if fermionic modes dominate. Supersymmetric vacua have
zero cosmological constant, due to bose-fermi cancellations.

The cosmological constant problem arises because any reasonable choice
for the ultraviolet cutoff scale $U$ leads to a $\Lambda$ which
exceeds the observational upper bound by a ridiculously large
multiple, roughly $10^{-100}$ (here I am invoking ``roughly'' as
per standard usage in cosmology, meaning order of magnitude in the
exponent). This is because particle physics scales such as the
Planck mass ($10^{19}$ GeV), the Standard Model Higgs vev
($10^2$ GeV), and the apparent supersymmetry-breaking scale
($10^2 - 10^3$ GeV), greatly exceed the energy scale characterizing the
current matter density of the universe ($10^{-3}$ eV).
This is a disturbing problem, made worse by our desire to allow
a rather large effective $\Lambda$ during an earlier inflationary
epoch. It should also be noted that recent ideas about quintessence
in no way address the main cosmological constant problem; rather
quintessence models evolve one very small effective $\Lambda$ value
to another very small effective $\Lambda$ value.

String theory, which is (if nothing else) a consistent theory of
quantum gravity, ought to give us some profound insight to this
problem. Unfortunately even with recent advances of the duality
revolution, the cosmological constant problem remains a complete
mystery even in string theory.

A possible ray of hope is provided by a suggestion of Witten\cite{witcc}, 
which ties
in nicely\cite{kachru} to some recent work on the idea of ``string islands''.
Witten observed that in 2+1 spacetime dimensions you can have supersymmetry
of the vacuum (and thus $\Lambda$$=$$0$) without supersymmetry of the
spectrum (i.e. no bose-fermi degeneracy for particles).
Furthermore, string theory in 2+1 dimensions actually becomes
string theory in 3+1 dimensions in the limit where the string coupling
(determined by the dilaton vev) goes to infinity. This peculiar
phenomenon is similar to that which leads to the 11-dimensional
M theory limit. The strongly-coupled 2+1 dimensional string theory
has light solitons, which actually behave exactly like a set of
light Kaluza-Klein modes. These solitonic degrees of freedom thus
represent the degrees of freedom of a third spatial dimension
compactified on a circle. In the strong coupling limit the
radius of this circle becomes infinite, and a 3+1 dimensional theory
results.

This suggests a method for finding non-supersymmetric string
vacua with zero cosmological constant, by starting with 2+1
dimensional string vacua which contain a dilaton. Note that it is
important for this trick that the 2+1 string vacua do not contain
any geometrical moduli associated with compactifications from
higher dimensions. Such moduli would invalidate the original
argument, leading presumably to a nonzero $\Lambda$ whose scale
is set by the square of the bose-fermi mass splittings divided by
the compactification scale; this is too large unless we manage to
keep all the mass splittings below about 100 GeV.

Thus the pure version of this trick requires string islands (peninsula?):
string vacua which contain the dilaton and its axion partner, but do
not contain any geometrical moduli. Surprisingly, such string islands
are known to exist even in the weakly-coupled limit of the heterotic
string\cite{chl}.
There is, for example, a 3+1 dimensional string vacuum whose low
energy limit is pure 3+1 dimensional $N=4$ supergravity.
Many more examples have been constructed recently\cite{harvey98}.

I should emphasize that even if one could exhibit string islands
corresponding to non-supersymmetric vacua with zero cosmological
constant, it is another matter entirely to show that any such vacuum
is consistent with the Standard Model. An important conclusion for
cosmologists, however, is that current thinking about making
string vacua which are more ``realistic'' seems to favor reduced
sets of moduli.
A broader conclusion is that in the long run cosmological considerations
are likely to play an important role in resolving mysteries about
the vacuum state of string theory.

\section*{Delightful D branes}

As mentioned above, string theory abounds with membranes of various types
and dimensionalities. Of particular interest are D branes, objects which,
considered as backgrounds for string propagation, preserve part of the
underlying spacetime supersymmetry. D branes have a number of special
properties, and occur with various dimensionalities (thus we have D
instantons, D particles, D strings, and D membranes with up to 9 spatial
dimensions). From the string point of view a D brane is a soliton whose
mass (or tension, or mass per unit volume) is proportional to the inverse
of the string coupling $g_s$. This means, among other things, that D branes
become light in those regions of string moduli space where the string
coupling is large.

This simple fact has led to a new interpretation for singularities
of various fixed spacetime backgrounds in which strings propagate. 
These singularities are associated with compactifying some of the original
9 or 10 spatial dimensions onto orbifolds, conifolds, or other singular
geometries. A D brane can wrap around a D-dimensional closed cycle of this
compact space; when this cycle is shrunk to a point a singularity appears,
associated with the vanishing mass of the wrapped D brane. 
This observation provides a generic and physically intuitive mechanism
for ``smoothing'' spacetime singularities.

The obvious question, of course, is whether this D brane smoothing also
applies to cosmological singularities.
Recent work suggests that the answer is yes\cite{wilczek,riotto}, although
perhaps not in all cases\cite{banks98}.
This is an exciting avenue for future research.

\section*{Dark Matter in Parallel Universes}

The mass scale at which string physics becomes stringy is known
as the string scale, $m_s$. For the weakly coupled heterotic string,
the string scale can be shown to be about $10^{18}$ GeV, only about
an order of magnitude smaller than the Planck mass, $m_p$.
However in other regions of string moduli space the string scale can
be much smaller\cite{witscale}. 
Since we don't know where we are in moduli space,
we also don't know the value of $m_s$. The most we can say at present
is that $m_s$ is greater than about 1 TeV, due to nonobservation of
stringy effects in the Tevatron collider experiments\cite{lykscale}.

It is tempting to imagine\cite{lykscale} 
that perhaps the string scale is not
too far above the current lower bound, in the multi-TeV region
which will eventually be accessible to colliders. If this bold hypothesis
is correct, we are immediately faced with the problem of explaining
the small ratio $m_s/m_p$. In string theory this small ratio is
presumably related to certain moduli having very large or very small
vevs, as measured in units of $m_s$. If these moduli are ``invisible''
moduli of the type discussed earlier, then their existence may have no
other direct consequences for observable low energy physics. 

On the other
hand, this small ratio could be a consequence of large compactified
dimensions\cite{dim1,dim2}, 
through a scaling relation like
\begin{equation}
m_p^2 \sim m_s^{n+2}\; R^n \quad ,
\end{equation}
where $R$ is the size of the large compact dimensions, and $n$ is the
number of such dimensions. For $m_s$ of order a TeV and
$n\ge 2$, $R$ in the above
relation can be as large as 1 mm!

Actually, this form of the large extra dimensions 
scenario is completely ruled
out by particle physics constraints, unless we make an additional
bold hypothesis: that the entire Standard 
Model gauge theory is confined to live on a membrane orthogonal
to the large extra dimensions. Since D branes are known to have
supersymmetric gauge theories confined to their worldvolumes, this
hypothesis fits rather nicely with our current picture of string
theory. If correct, the graviton has many massive Kaluza-Klein
copies, but the Standard Model particles know of the existence
of large extra dimensions only through coupling to gravity.
There are, not surprisingly, many interesting cosmological implications
of this scenario\cite{dim2,dim3,linde98}.

One intriguing observation is that, if the Standard Model gauge theory
is confined to some configuration of branes, then there may be
other gauge theories confined to other brane configurations, separated
from us in one or more of the large extra dimensions. Such hidden
sectors are very much like parallel universes, except that they
are gravitationally coupled to the visible universe. If these other
``brane-worlds'' contain stable matter, planets, stars, galaxies, etc.,
these will all appear to us as dark matter. Since the
laws of (non-gravitational) physics could be quite different in these
parallel worlds, qualitatively new forms of macroscopic matter may also
be produced. 
It would be interesting to determine the current observational bounds
on (i) dark ``planets'' in the vicinity of our solar system, (ii) dark
``stars'' within our galaxy and the galactic halo, and (iii) the density
and distribution of dark ``galaxies''.

\end{document}